\documentclass[prl,aps, superscriptaddress,twocolumn]{revtex4-1}
\usepackage{amsmath,amssymb}
\usepackage{graphicx}
\usepackage{wasysym}
\usepackage{amsfonts}
\usepackage{bm}
\usepackage{enumerate}
\usepackage{color}
\usepackage[resetlabels]{multibib}
\usepackage{epstopdf}
\usepackage{latexsym}
\usepackage[breaklinks,colorlinks = true,linkcolor = red,urlcolor=cyan,citecolor=red]{hyperref}
\usepackage[caption=false,singlelinecheck=false]{subfig}

\usepackage{times}
\newcommand{\bea}{\begin{eqnarray}}
\newcommand{\eea}{\end{eqnarray}}
\newcommand{\be}{\begin{eqnarray}}
\newcommand{\ee}{\end{eqnarray}}
\newcommand{\bw}{\begin{widetext}}
\newcommand{\ew}{\end{widetext}}

\begin{document}
\title{Spin-helix driven insulating phase in two dimensional lattice}
\author{Hyun-Jun K. Park}
\affiliation{Department of Physics, Korea Advanced Institute of Science and Technology, Daejeon, 34141, Korea}
\author{Hyeok-Jun Yang}
\email{yang267814@kaist.ac.kr}
\affiliation{Department of Physics, Korea Advanced Institute of Science and Technology, Daejeon, 34141, Korea}
\author{SungBin Lee}
\email{sungbin@kaist.ac.kr}
\affiliation{Department of Physics, Korea Advanced Institute of Science and Technology, Daejeon, 34141, Korea}
\date{\today}
\begin{abstract}
Motivated by emergent $SU(2)$ symmetry in the spin orbit coupled system, we study the spin helix driven insulating phase in two dimensional lattice. When both Rashba and Dresselhaus spin orbit couplings are present, the perfect Fermi surface nesting occurs at a special condition depending on the lattice geometry. In this case, the energies of spin up at any wave vector $\vec{k}$ are equivalent to the ones of spin down at $\vec{k}\!+\!\vec{Q}$ with so-called the \textit{shifting wave vector} $\vec{Q}$. Thus, the system stabilizes magnetic insulator with spiral like magnetic ordering even in the presence of tiny electron-electron interaction where the magnetic ordering wave vector is proportional to $\vec{Q}$.  
We first show the condition for existence of the \textit{shifting wave vector} in general lattice model and emergent $SU(2)$ symmetry in the spin orbit coupled system. 
Then, we exemplify this in square lattice at half filling and discuss the insulating phase with (non-) coplanar spin density wave and charge order. Our study emphasizes possible new types of two dimensional magnetic materials and can be applicable to various van-der Waals materials and their heterostructures with the control of electric field, strain and pressure. 
\end{abstract}


\maketitle
\section{Introduction}
Spin $SU(2)$ symmetry, invariance with respect to electron spin rotation, is an important quantity giving conservation of spin polarization in the system. When spin and orbital degrees of freedom are coupled, however, such $SU(2)$ symmetry is generally broken thus spin polarization becomes no longer a good quantum number. Particularly for two dimensional system including surfaces, two different types of spin orbit couplings (SOC) are mainly addressed due to broken inversion symmetries, Rashba and Dresselhaus spin orbit couplings. The Rashba effect originates from the effective electric field at the surface or interface of crystal structures, whereas, the Dresselhaus effect comes from the bulk inversion asymmetry {\sl i.e.} $\epsilon_{\uparrow} ({\vec{k}}) \!\neq\! \epsilon_{\downarrow} ({-\vec{k}})$. 
Their measurement and controllability have been widely studied in quantum wells  for several decades\cite{Kor09}.
More recently, the possible creation of such interactions has been explored even in optical lattices\cite{Lin11,Gru17}.

Despite general consensus that spin orbit effect breaks spin $SU(2)$ symmetry, some exotic cases are known to host emergent symmetry higher than $SU(2)$ such as $SU(4)$ as a result of strong spin and orbital coupling\cite{Yam18}.
Moreover, it has been pointed out that well control of Rashba and Dresselhaus effects can give rise to emergent $SU(2)$ symmetry. In particular, when Rashba and Dresselhaus spin orbit couplings are equal in two dimensional electron gas, it can be exactly mapped into the Dresselhaus [110] model which describes quantum well grown along [110] direction\cite{Ohn99,Ali10}. In this case, the effective magnetic field via spin orbit coupling is unidirectionally induced,  thus spin polarization of a helical mode is conserved along that special direction. This leads to a new type of $SU(2)$ symmetry in the system based on the transformed spin operators with the \textit{shifting wave vector} where magnitude is proportional to the strength of spin orbit coupling. Such emergent $SU(2)$ symmetry with persistent spin helix has become one of the central topics of experimental realization in semiconductor quantum wells and their direct observations\cite{Lin11,Gru17,Wal12,Det17}.

In general, microscopic control of spin orbit coupling has been regarded as very challenging tasks.  
In two dimensional electron gas of semiconductors, however, possible control of spin orbit coupling is studied via gate electric field or tunneling current and etc\cite{Nit97,Eng97,Kog02,Che18}. 
Moreover, at the interface of oxide heterostructures and van der waals heterostructures, the realization of a giant spin orbit coupling with heavy ions is extensively studied both in experimental and theoretical aspects\cite{Kim16,Man15,Ish11,Zho15,Cav10}. Such controllability and enhancement of spin orbit coupling have gotten much attention pursuing control of spin precession potentially related to spintronic devices and quantum computations.  
Along with controlling SOCs mentioned above, interplay of electron correlation can also give rise to new emergent phenomena. 
Especially in oxide heterostructures and transition metal chalcogenide heterostructures, both SOC and electron correlation play an important role. 

In this paper, taking into account possible controllability of SOC in two dimensional heterostructures, we discuss the unique magnetic insulator induced by interplay of SOC and electron correlation. Considering both Rashba and Dresselhaus SOCs and their special ratio, we show the presence of  a \textit{shifting wave vector} which depends on the lattice geometry. The \textit{shifting wave vector} satisfies $\epsilon_{\uparrow}({\vec k + \vec Q }) = \epsilon_{\downarrow} ({\vec k})$ where $\vec{Q}$ is proportional to the magnitude of both SOCs. In this situation, any infinitesimal electron interaction gives rise to perfect Fermi surface nesting and results in stabilization of magnetically ordered insulator where the ordering wave vector is proportional to the \textit{shifting wave vector} $\vec{Q}$.

We exemplify our scenario in the case of simple square lattice at half filling. 
As a result, we find the insulating phase stabilized by non-coplanar spin density wave in addition to charge order, which are characterized by the \textit{shifting wave vector} discussed above. 
By controlling relative ratio between Rashba and Dresselhaus SOCs, we discuss the possible phase diagram for metal-insulator transition and generalization to other lattice geometry. Our work opens a new mechanism of two dimensional magnetic insulators applicable to many transition metal oxides and chalcogenides, and their heterostructures.

%

{\bf \em General Lattice Model and Shifting Wave Vector ---}
On a two-dimensional lattice, we consider the Hubbard model with SOCs as
\begin{align}
H_{t,U}&=-\sum_{i,j,\sigma}t_{ij}c_{i,\sigma}^{\dagger}c_{j,\sigma}+h.c. + U\sum_{i}n_{i\uparrow}n_{i\downarrow}\label{eqn:hamtU}\\
H_{soc}&=i\sum_{i,j,\sigma}s_{ij}(\hat{d}_{ij}\cdot\vec{\sigma})c_{i,\sigma}^{\dagger}c_{j,\sigma}+h.c. \label{eqn:hams},
\end{align}
where $c_{i,\sigma}^\dagger$, $c_{i,\sigma}$ and $n_{i\sigma}$ are electron  creation, annihilation and number operators at site $i$ with spin $\sigma \!\in\! (\uparrow,\downarrow)$  respectively. 
In Eqs.\eqref{eqn:hamtU} and \eqref{eqn:hams}, the lattice sites $i,j$ are not restricted to the nearest neighbors and the unit vector $\hat{d}_{ij}$ characterizes the SOC with the Pauli matrices $\vec{\sigma}\!=\!(\sigma^x,\sigma^y,\sigma^z)$. 

Let's first consider the non-interacting system ($U=0$) and examine the Fermi surface properties. We focus on the case when the magnitudes of Rashba and Dresselhaus SOCs satisfy a particular ratio specific to the lattice geometry. (We will show this later.)  Then, the SOC in Eq.\eqref{eqn:hams} after a proper spin rotation can be written as 
\bea
\tilde{H}_{soc}=i\sigma^z \sum_{i,j,\sigma}s_{ij}(c_{i,\sigma}^{\dagger}c_{j,\sigma}-c_{j,\sigma}^{\dagger}c_{i,\sigma}).
\label{eqn:hams_m}
\eea
In this limit, our aim is to convince the existence of the \textit{shifting wave vector} $\vec{Q}$ which is defined to satisfy the shifting condition, $\epsilon_{\vec{k}+\vec{Q},\uparrow}\!=\! \epsilon_{\vec{k},\downarrow}$ where $\epsilon_{\vec{k},\sigma^z}$ is the energy dispersion in $\vec{k}$-space with spin $\uparrow$, $\downarrow$.
For a given hopping magnitude $t_{ij}$ and the SOC strength $s_{ij}$, 
the \textit{shifting wave vector} $\vec{Q}$ always exists if they meets
\begin{equation}
s_{ij}=t_{ij}\tan (\vec{Q}\cdot (\vec{x}_i-\vec{x}_j)/2).
\label{eqn:stconn}
\end{equation}
This indicates that it is possible to choose the appropriate SOC strength $s_{ij}$ for an arbitrary vector $\vec{Q}$ and the hopping strength $t_{ij}$.  Generically, it can be shown that a two-level Hamiltonian with a form $H_{\sigma^z}(k)=h(k) {I_{2\times 2}}+ g(k) \sigma^z$  admits this shifting conditions. (See Supplementary Information.)
To study the nesting property, we define the spin operators
\begin{gather}
S_{\vec{k},\vec{Q}}^-  = c_{\vec{k},\downarrow}^\dagger c_{\vec{k}+\vec{Q},\uparrow},\quad
S_{\vec{k},\vec{Q}}^+ = c_{\vec{k}+\vec{Q},\uparrow}^\dagger c_{\vec{k},\downarrow},\nonumber\\[1ex]
S_{\vec{k},0}^z = c_{\vec{k},\uparrow}^\dagger c_{\vec{k},\uparrow} - c_{\vec{k},\downarrow}^\dagger c_{\vec{k},\downarrow},
\label{eqn:SpinOpe}
\end{gather}
obeying the standard $SU(2)$ commutation relations $[S_{\vec{k},0}^z, S_{\vec{k},\vec{Q}}^\pm] 
\!=\! \pm 2 S_{\vec{k},\vec{Q}}^\pm$ $[S_{\vec{k},\vec{Q}}^+ , S_{\vec{k},\vec{Q}}^-]\!=\! S_{\vec{k},0}^z$. Since they commute with the Hamiltonian  $[H, S_{\vec{k},\vec{Q}}^\pm]\!=\![H, S_{\vec{k},0}^z]\!=\!0$, the emergent $SU(2)$ symmetry enables us to manipulate Eq.(\ref{eqn:SpinOpe}) for the magnetic order parameters.

{\bf \em Perfect Nesting in Half-Filled Square Lattice --- }
Now suppose the square lattice and the nearest neighbor hoppings $t_{ij}=t$ only. The shifting condition Eq.(\ref{eqn:stconn}) requires $s_{ij}=\pm s_{x (y)}$ for $\vec{r}_i=\vec{r}_j\pm \hat{x} (\hat{y})$ and $s_{ij}=0$ otherwise. We can choose $(s_x,s_y)=t(\tan(Q_x/2),\tan(Q_y/2))$ for a given shifting vector $\vec{Q}=(Q_x,Q_y)$. Inversely, the shifting vector $\vec{Q}={2}(\tan^{-1}(s_x/t),\tan^{-1}(s_y/t))$ can be designated for arbitrary strength $(s_x,s_y)$. With the nearest neighbor SOCs, 
Eq.(\ref{eqn:hams}) is decomposed into Rashba and Dresselhaus SOCs,
\begin{equation}
\begin{split}
H_R&= is_R\sum_{\vec{r}}\Big(-\sigma^y_{\alpha \beta} c_{\vec{r},\alpha}^\dagger c_{\vec{r}+\hat{y},\beta}+\sigma^x_{\alpha\beta}c_{\vec{r},\alpha}^\dagger c_{\vec{r}+\hat{x},\beta}\Big)+\text{h.c}.
\\
H_D&= is_D\sum_{\vec{r}}\Big(\sigma^x_{\alpha\beta}c^{\dagger}_{\vec{r},\alpha}c_{\vec{r}+\hat{x},\beta}-\sigma^y_{\alpha\beta}c^{\dagger}_{\vec{r},\alpha}c_{\vec{r}+\hat{y},\beta} \Big)+\text{h.c}.
\end{split}
\label{eq:SOCrd}
\end{equation} 
The prerequisite for the shifting condition Eq.(\ref{eqn:hams_m}) can be attained
when $s_R=s_D\equiv s/\sqrt{2}$ whose ratio $s_R/s_D=1$ is specific to the square lattice. 
After $\pi/2$ spin rotation along ($\hat{x}\!+\!\hat{y}$) axis, $H_R+H_D$ is rearranged into the form of Eq.\eqref{eqn:hams_m} with the dispersion $\epsilon_{\vec{k},\sigma^{z}}=-2t\left(\cos(k_x)+\cos(k_y)\right)
 +2s\sigma^{z}\left(\sin(k_x)+\sin(k_y)\right)
$. It satisfies the shifting property $\epsilon_{\vec{k}+\vec{Q},\uparrow}=\epsilon_{\vec{k},\downarrow}$ in the rotated spin basis, with the shifting  wave vector $\vec{Q}={2\tan^{-1}(s/t)}(1,1)$.
For a small SOC, this agrees with the results of free electron gas since the hopping amplitude is inversely proportional to the effective mass.\cite{Ber04}

\begin{figure}{}
    \centering
    \subfloat[]{\label{fig:fer}\includegraphics[width=0.5\linewidth]{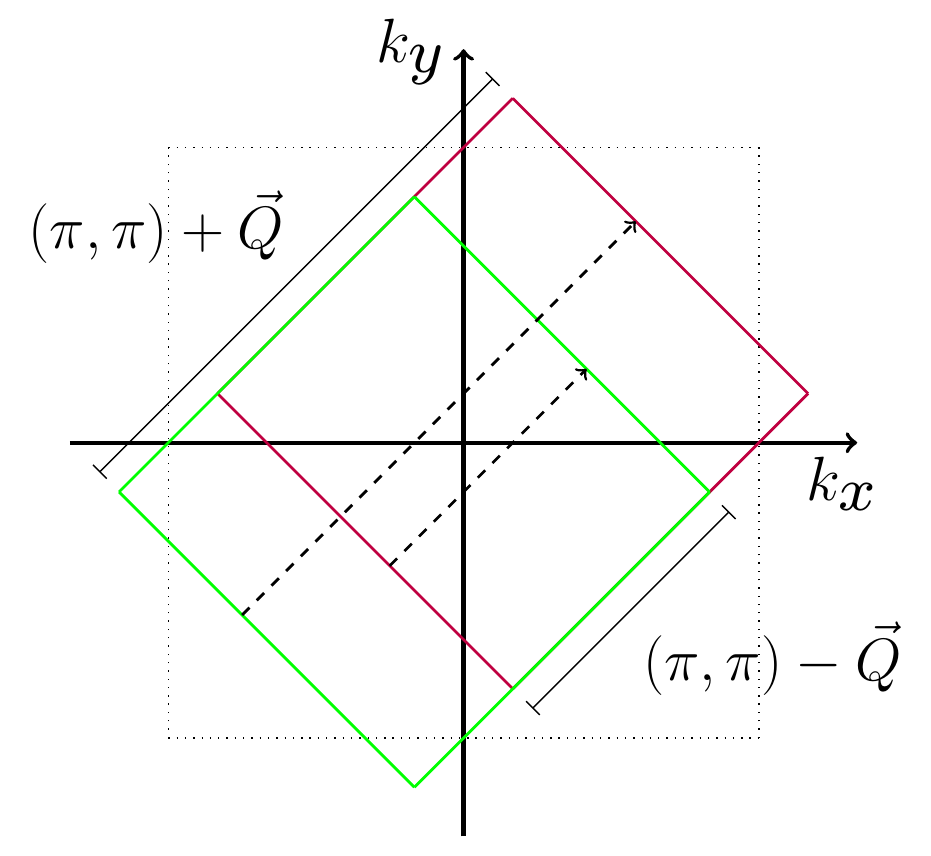}}
    \subfloat[]{\label{fig:sus}\includegraphics[width=0.5\linewidth]{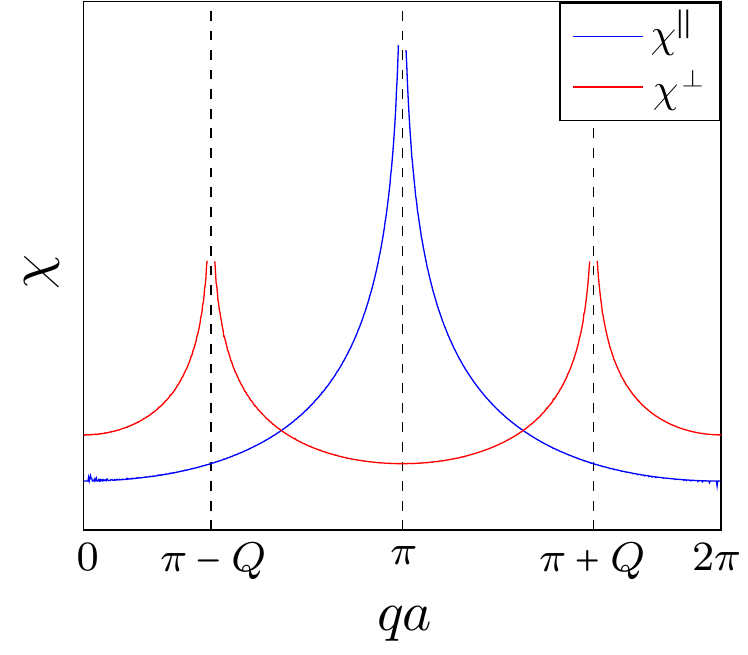}}
    \caption{Fermi surface and spin susceptibilities for square lattice at half-filing with equal magnitudes of Rashba and Dresselhaus spin orbit couplings. (a) The splitted Fermi surfaces with the shifting vector $\vec{Q}=(0.6\pi, 0.6\pi)$. With a proper spin rotation, the green and purple color represent the Fermi surfaces for up and down-spin electrons respectively. The dashed lines correspond to the wave vectors $\vec{K}_\pm$ for perfect Fermi surface nesting. 
   (b) Spin susceptibilities $\chi^{||}(\vec{q})$ and $\chi^{\perp}(\vec{q})$ for $\vec{q}=(q,q)$. }   
  \end{figure}

Taking into account the shifting property in case $s_R=s_D$, the Fermi surfaces for up/down spins are perfectly split as shown in Fig. \ref{fig:fer}. This generates the nesting between Fermi surfaces for opposite spins with the wave vectors $\vec{K}_{\pm}=(\pi,\pi)\pm\vec{Q}$.
Furthermore, due to the shape of the Fermi surface itself at half filling, there is additional nesting for each Fermi surface with the wave vector $\vec{K}^*=(\pi,\pi)$. 
Having these in mind, one expects the magnetic instabilities near $\vec{K}_{\pm}$ and $\vec{K}^{*}$.
The spin susceptibility at a wave vector $\vec{q}$ is related to the scattering amplitude between the filled and empty states near Fermi level separated by $\vec{q}$, 
$\chi^\mu(\vec{q})\!=\!-\!\sum_{\vec{k},\sigma} \big( {f(\epsilon_{\vec{k},\sigma})\!-\!f(\epsilon_{\vec{k}+\vec{q},\sigma'})}\big)/ \big({\epsilon_{\vec{k},\sigma} \!-\! \epsilon_{\vec{k}+\vec{q},\sigma'}}\big)$. ($\mu=\perp,||$)\cite{Faz99} Here, $\sigma \!=\! -\sigma'$ for $\chi^\perp$, whereas $\sigma \!=\! \sigma'$ for $\chi^{||}$ and $f(\epsilon_{\vec{k},\sigma})$ is the occupation number of the energy $\epsilon_{\vec{k},\sigma}$. It is noteworthy that the instability of $\chi^{||}(\vec{q})$ near $\vec{K}^*$ is special for square lattice at half filling, while the behavior of
$\chi^{\perp}(\vec{q})$ near $\vec{K}_\pm$ is comprehensive regardless of the lattice geometry and the electronic filling.

Along the line $k_x\!=\!k_y$ in $\vec{k}$-space, the spin susceptibility $\chi^{||}(q, q)$ is approximately evaluated as $ \chi^{||}(q, q)\thicksim F(q)$ where $
F(q) = \frac{1}{\text{sin}(q/2)}\text{log}\Big[\frac{1+\text{sin}(q/2)}{1-\text{sin}(q/2)}\Big]
$. Divergence of $F(x)$ at $x=\pi$ represents the magnetic instability at $\vec{q}=\vec{K}^*$.
Similarly, one can estimate $\chi^{\perp}(q, q)$ as $\chi^{\perp}(q, q) \thicksim F(q+Q)$ for $0<q<\pi$, $ \chi^{\perp}(q, q) \thicksim F(q-Q)$ for $ \pi<q<2\pi$. This implies that the perfect Fermi surface nestings at $\vec{q}=\vec{K}^{*}$ and $\vec{K}_{\pm}$ are associated with the spontaneous magnetization $\langle S^{||}_{\vec{K}^{*}}\rangle$ and $\langle S^{\perp}_{\vec{K}_{\pm}}\rangle$ respectively.  As the magnitude of SOC, $s$, varies, the peak of $\chi^\perp(\vec{q})$ moves with the shifting vector $\vec{Q}$ while $\chi^{||}(\vec{q})$ doesn't. In Fig.\ref{fig:sus}, the spin susceptibilites are numerically evaluated with $Q=0.6\pi$ as functions of $\vec{q}=(q,q)$. Indeed, the magnetic susceptibilities $\chi^\perp(\vec{q})$ and $\chi^{||}(\vec{q})$ diverge at $K_{\pm}$ and $K^*$ respectively. The overall weight near the peak reflects the density of states in which the nesting happens close to the Fermi surfaces. The magnetic susceptibility $\vert\chi^{||}(\vec{q}\thickapprox\vec{K}^*)\vert$ originates from the density of states for both spins, while $\vert\chi^\perp(\vec{q}\thickapprox\vec{K}_{\pm})\vert$ is proportional to the ones for each spin, which explains half of the magnitude compared to $\vert\chi^{||}(\vec{q}\thickapprox\vec{K}^*)\vert$. 

{\bf \em Magnetic insulator with spin-helix ordering --- }
With the magnetic instabilities at the wavevectors $\vec{K}_{\pm}$, $\vec{K}^*$,  the ground state is expected to stabilize the spin and charge orderings with electron interaction $U$. We consider the quartic term $Un_{i\uparrow}n_{i\downarrow}$ with the Hartree-Fock approximation, keeping the dominant orderings $\langle S^{\perp}_{\vec{K}_{\pm}}\rangle $, $\langle S^{||}_{\vec{K}^*} \rangle$ and $\langle n_{\vec{K}^{*}}\rangle$ only.
We further generalize our analysis when the ratio of Rashba SOC to Dresselhaus SOC deviates from emergent $SU(2)$ symmetric point, $\delta\equiv s_R-s_D \neq 0$.
  A small but finite $\delta$ implies the collapse of the perfect Fermi surface nesting. The polarized spin identifying the band dispersion is now momentum dependent, $\epsilon_{\vec{k},\sigma^z}\rightarrow \epsilon_{\vec{k},\tilde{\sigma}^z(\theta_{\vec{k}})}$. (See Supplementary Information for details.)
Unlike the case of perfect Fermi surface nesting, the critical value of electron interaction $U$ is required to stabilized the magnetic order when $\delta \neq 0$. Similar to the Stoner criterion for a ferromagnet, the critical values of $U$ are inversely proportional to the maximum peak of the magnetic susceptibilities $\chi_{\delta}^{\perp}(\vec{K}_{\pm})$ and $\chi_{\delta}^{||}(\vec{K}^*)$, which are damped for $\delta \neq 0$. 

\begin{figure}[t]
    \centering
    \includegraphics[width=0.8\linewidth]{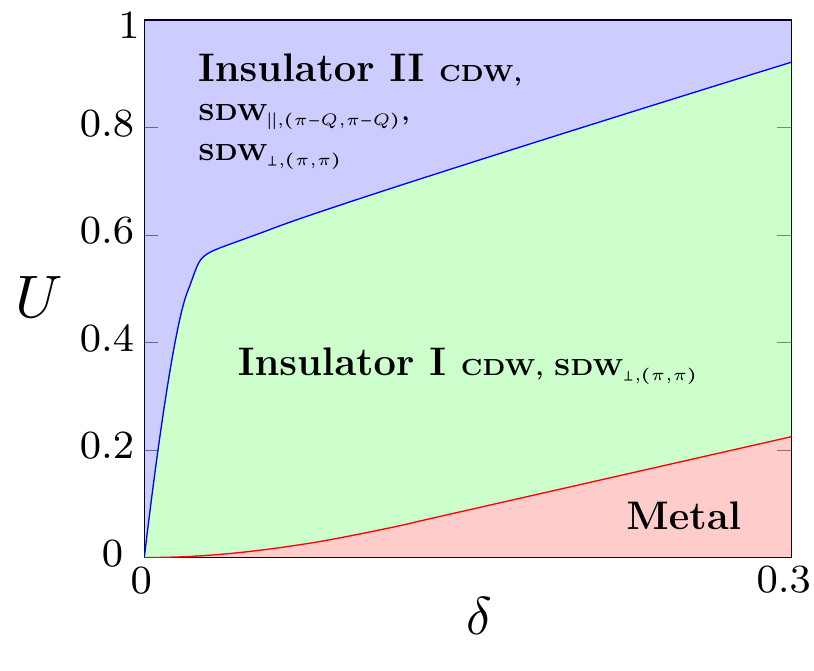}
    \caption{The phase diagram as functions of electron interaction $U$ and $\delta=s_R-s_D$. Insulator I : charge density wave and spin density wave exist with the wave vector $\vec{K}^*=(\pi,\pi)$ where the collinear magnetic moments are (anti-) parallel to $\hat{x}+\hat{y}$ direction. Insulator II : In addition to charge and spin density wave of Insulator I, spiral magnetic order with the wave vector $\vec{K}_\pm = (\pi \pm Q,\pi \pm Q)$ is further stabilized on the plane perpendicular to $\hat{x}+\hat{y}$, resulting in non-coplanar spin states. See Fig.\ref{fig:cha} for details of charge and spin density waves.
    }
    \label{fig:dsuscep}
\end{figure}

\begin{figure}[t]
    \centering
    \includegraphics[width=0.8\linewidth]{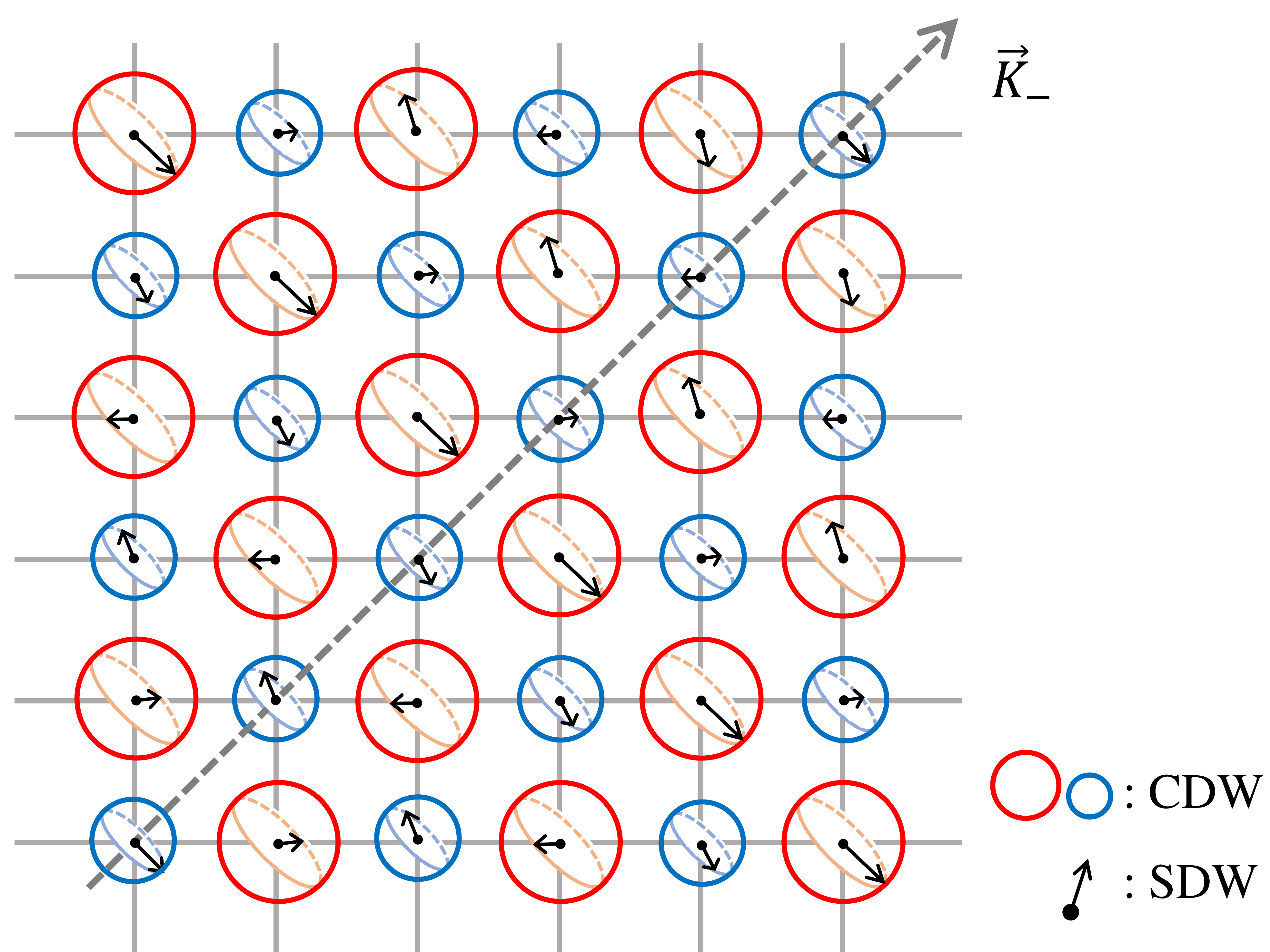}
    \caption{Schematic depiction of the insulator II phase with $\vec{Q}=(0.6\pi, 0.6\pi)$. The circle size and caged arrow represent the amount of charge order $\langle n_{\vec{K}^{*}}\rangle$ and (in-plane perpendicular $\hat{x}+\hat{y}$ axis) magnetic order $\langle S^{\perp}_{\vec{K}_{-}}\rangle $ respectively. The out-of plane magnetization  $\langle S^{||}_{\vec{K}^{*}}\rangle$, which is (anti-) parallel to $\hat{x}+\hat{y}$ axis, is built in the (red) blue circle color with magnitude proportional to the circle size. 
}
    \label{fig:cha}
\end{figure}

Fig.\ref{fig:dsuscep} represents the phase diagram as functions of electron interaction $U$ and $\delta=s_R-s_D$. With increasing electron interaction $U$, there exists the phase transition from metal to insulator I phase where both spin density wave (SDW) and charge density wave (CDW) are stabilized with the wave vector $\vec{K}^*$. 
Further increasing electron interaction $U$, the system develops insulator II phase and stabilizes the SDW with the wave vector $\vec{K}^\pm$ in addition to CDW and SDW with the wave vector $\vec{K}^*$. Thus, in this regime, the system develops the non-coplanar magnetic ordering both with $\langle S^{\perp} (\vec{r})\rangle \sim \text{cos}(\vec{K}^{\pm} \cdot \vec{r})$ and $\langle S^{||} (\vec{r})\rangle \sim \text{cos}(\vec{K}^{*} \cdot \vec{r})$ in addition to the charge order. Fig.\ref{fig:cha}  shows the non-coplanar magnetic order along with charge order specifically with $\vec{K}^{-}=(0.4\pi, 0.4\pi)$ and $\vec{K}^*$. It is worth to note that the insulator I phase is specific to the square lattice at half-filling, while the generic metal-insulator phase transition will occur between the metal and the insulator II regime.

{\bf \em Discussion --- }
We have studied the possible spin-helix driven insulating phases, exemplifying the case of square lattice at half filling. In this case, the shifting vector exists  with  an emergent $SU(2)$ symmetry when Rashba and Dresselhaus SOCs take equal magnitudes. However, as mentioned above, our argument generally holds for other lattice geometries by controlling SOCs in different ratios. Thus, here we briefly discuss how to obtain emergent $SU(2)$ symmetry for the case of triangular lattice and compare with the case of square lattice. 

\begin{figure}[t]
\subfloat[]{\label{fig:SOC1}\includegraphics[width=0.23\textwidth]{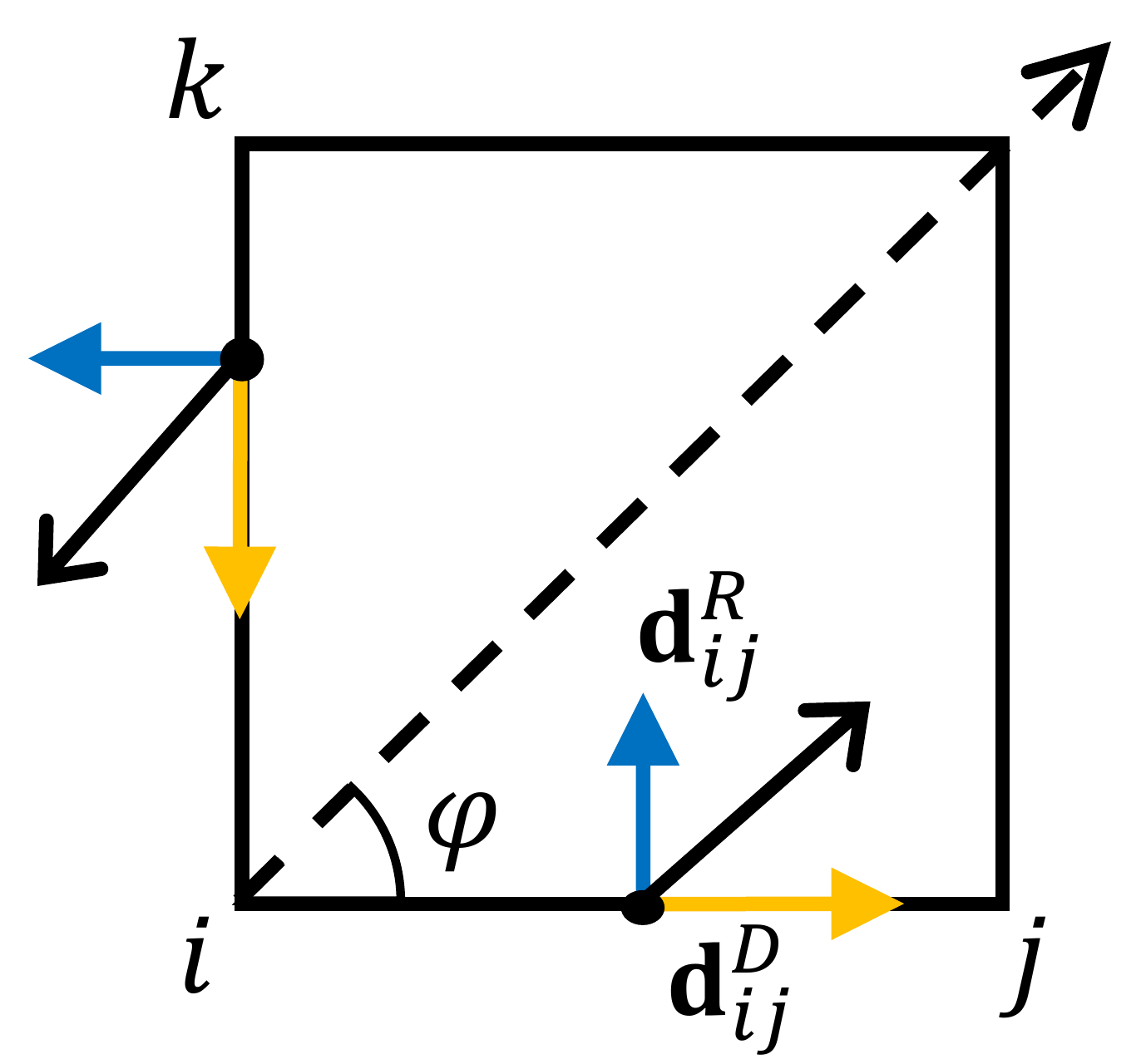}}
\subfloat[]{\label{fig:SOC2}\includegraphics[width=0.22\textwidth]{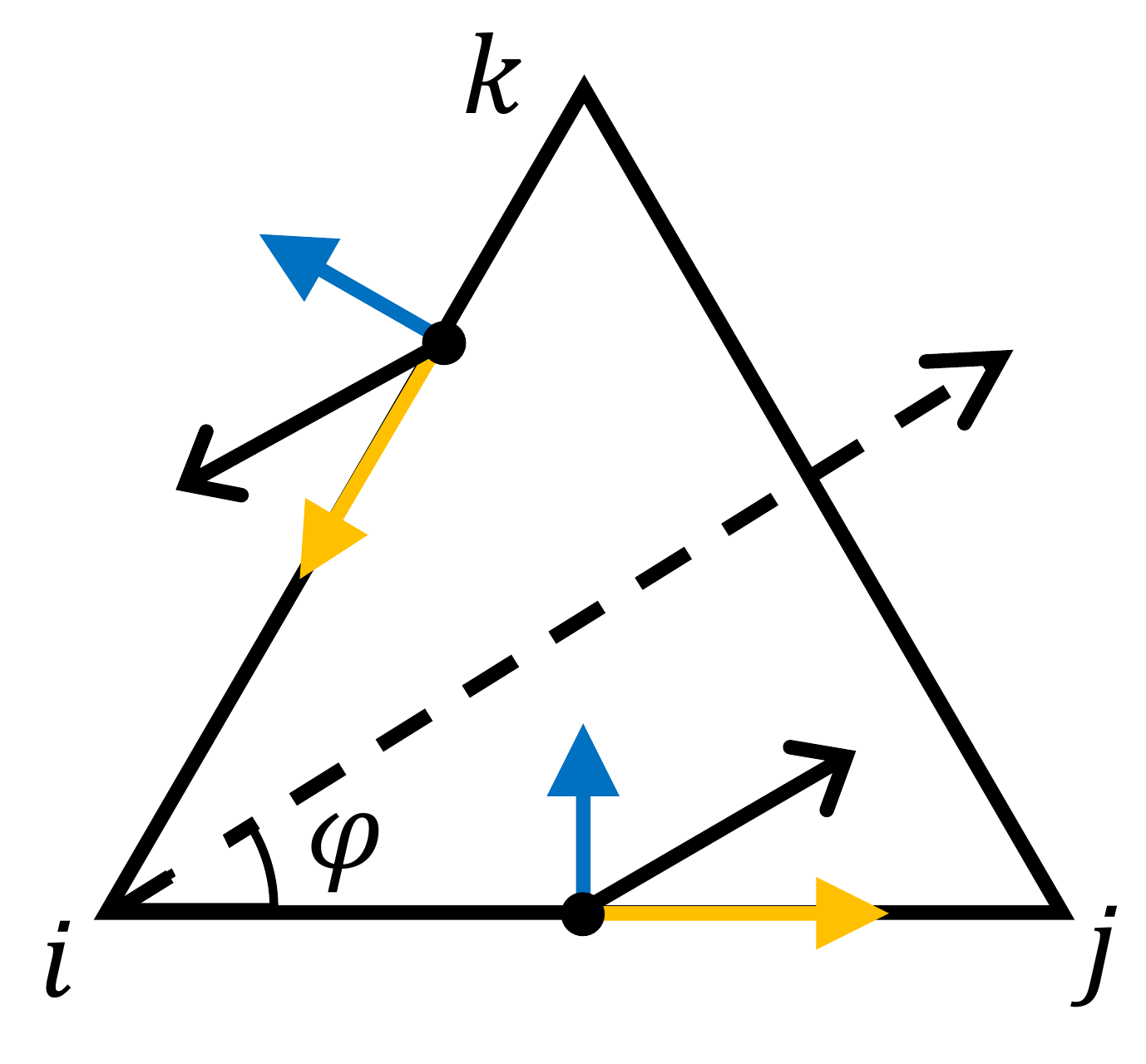}}
\caption{The characterizing SOC vector $s_{ij}\hat{d}_{ij}=\textbf{d}_{ij}^D+\textbf{d}_{ij}^R$ on the link $ij$ and the mirror reflected one $s_{ik}\hat{d}_{ik}$ about the mirror plane (dashed line) on the (a) square lattice and (b) triangular lattice. The SOC vector $s_{ij}\hat{d}_{ij}$ (black arrow) is decomposed as a sum of the Dresselhaus (yellow arrow) and Rashba SOCs (blue arrow).
}
\label{fig:SOC condition}
\end{figure}

In order to obtain the emergent $SU(2)$ symmetry, combinations of SOCs should be the form of Eq.\eqref{eqn:hams_m} after a proper spin rotation. 
Eq.(\ref{eqn:hams_m}) can be constructed when the characterizing vector $\hat{d}_{ij}$ in Eq.(\ref{eqn:hams}) is aligned in a collinear fashion. 
The characterizing vector $s_{ij}\hat{d}_{ij}$ in Eq.\eqref{eqn:hams} is decomposed as a sum of the Dresselhaus and Rashba SOCs, $s_{ij}\hat{d}_{ij}=\textbf{d}_{ij}^D+\textbf{d}_{ij}^R=d_{ij}^D\hat{x}+d_{ij}^R\hat{y}$ in the coordinate whose $\hat{x}$-axis towards the link $ij$. 
In the presence of the translational invariance, all vectors $s_{ij}\hat{d}_{ij}$ on other links can be obtained by the lattice translations combined with the mirror reflection. Here, for simplicity, we consider the mirror planes (containing dashed lines in Fig.\ref{fig:SOC condition}), which bisect the acute angles between two adjacent links. 
 Given the vector $\hat{d}_{ij}$ on the link $ij$, the adjacent vector $\hat{d}_{ik}$ is obtained by implementing the mirror reflection about the bisecting plane. Thus, the vectors $\hat{d}_{ij}$ and $\hat{d}_{ik}$ are collinear if and only if $\hat{d}_{ij}$ is parallel or perpendicular to the mirror plane, satisfying 
$
{d_{ij}^R}/{d_{ij}^D}\!=\!\text{tan}(\varphi)$ or ${d_{ij}^R}/{d_{ij}^D}\!=\!\text{tan}(\pi/2\!+\!\varphi)$ 
where $\varphi$ is the angle between the mirror plane and the lattice link. This angle corresponds to the half acute angle between two adjacent links emanated from $i$ to $j$ and $k$. 
For example, the square lattice with $\varphi=\pi/4$ demands $\vert\textbf{d}_{ij}^{R}\vert=\vert\textbf{d}_{ij}^{D}\vert$ on every link. 

Now, we move on to the triangular lattice. Unlike the square lattice, there is a link $jk$ perpendicular to the mirror plane, which allows only $\textbf{d}_{jk}^{D}$ thus $\textbf{d}_{jk}^{R}=0$. 
This link keeps arbitrary magnitudes of $\textbf{d}_{jk}^{D}$ when $\hat{d}_{ij}$ is perpendicular to the mirror plane.
Then on the triangular lattice with $\varphi=\pi/6$, the SOC condition becomes $\text{d}_{ij}^{R}=\text{d}_{ij}^{D}/\sqrt{3}$ or $\text{d}_{ij}^{R}=-\sqrt{3}\text{d}_{ij}^{D}$. On the perpendicular link $jk$, $\textbf{d}_{jk}^{R}$ is forbidden in both cases while arbitrary $\textbf{d}_{jk}^{D}$ is allowed only in latter case $\text{d}_{ij}^{R}=-\sqrt{3}\text{d}_{ij}^{D}$.
In this way, one can satisfy the nesting condition with the emergent $SU(2)$ symmetry by controlling the SOCs for generic two-dimensional lattice.

Perfect Fermi surface nesting and emergent $SU(2)$ symmetry are present exactly at fine tuning of Rashba and Dresselhaus SOCs to satisfy special ratio, which is very challenging.
However, as we have discussed above, Fermi surface nesting induced by the combination of SOCs is still present at vicinity of such fine tuning even away from the exact ratio between SOCs. In principle, the effective electric fields for both in plane and out of plane can be controlled via applying gate voltage, pressure or strain effect due to substrate.
Thus, at surfaces or interfaces of heterostructures, one may be able to access parameter range where the system stabilizes spin helix driven insulating phase due to combination of SOCs. Our theoretical study opens a new way to approach unique magnetic insulators induced by SOC effect in two dimensional materials and their heterostructures. In future, it would be interesting to explore experimental controllability of SOCs in two dimensional materials especially transition metal chalcogenide series and oxide heterostructures and look for magnetic insulators and their phase transitions. 

\bibliography{biblio}

\newpage
\appendix*

\section{Appendix: Shifting property in General Function}
All the discussions above are done for the following sense. Suppose that we have the non-perturbed Hamiltonian with eigenvalue $f(x)$, where $x$ is some appropriate good quantum number; for the above cases, $x$ is the momentum. Now for some $a$, we want to find out the perturbating Hamiltonian with eigenvalue $g(x)$, in above the spin-orbit coupling term, which satisfies
\begin{equation}
f(x+a)-g(x+a)=f(x)+g(x).
\end{equation}
In this sense, we are going to call $f$ as the base function, $g$ as the shifting function, and $a$ as the shifting value. The equation above is then called the shifting property.

We have seen two problems which has the formation as above. In the original paper by Bernevig, we found out that for $f(x)=x^2$ and $g(x)=ax$ with $a$ be the shifting value gives the shifting property. In the research above,  We have the shifting property for $f(x)=\cos(x), g(x)=\tan(a/2)\sin(x)$, where $a$ is the shifting value. From these results, it is possible to ask the question that in which case are we possible to find out the shifting property. More precisely, we want some conditions for the base function $f$ such that for any shifting value $a$, there exists a shifting function $g$ which satisfies the shifting property.

Before find out the existence of such shifting function, we first discuss about the uniqueness. Suppose that there are two shifting functions $g,g'$ which satisfies the shifting properties. Then we have $f(x+a)-f(x)=g(x)+g(x+a)=g'(x)+g'(x+a)$, which satisfies $g(x)-g'(x)=-(g(x+a)-g'(x+a))$. Letting $h(x)=g(x)-g'(x)$ gives that $h(x)=-h(x+a)$. Notice that $h$ must be periodic with period $2a$. Such $h$ can be found easily, for example, $\sin(\pi x/a)$. Let $H_a=\{h(x)|h(x)=-h(x+a),\forall x\in \mathbb{R}\}.$ If two functions $g,g'$ satisfies the property $g(x)-g'(x)\in H_a$, then we say $g\sim_a g'$, which is the equivalence relation. Then we can see the following:

\textbf{Theorem.} For the base function $f$ and shifting value $a$, if we have the shifting function $g$, then it is unique under the equivalence relation $\sim_a$.

Hence if we find one shifting function, then we can find out the whole shifting function structure. Now we observe following properties:

\textbf{Properties.} Fix the shifting value $a$. Suppose the base functions $f,f'$ are differentiable and integrable. Then,
\begin{enumerate}
\item If $g, g'$ is the shifting function of $f,f'$ respectively, then $g+g'$ is the shifting function of $f+f'$.
\item $dg/dx$ is the shifting function of $df/dx$.
\item $G=\int gdx$ is the shifting function of $F=\int fdx$. Here, the integration constant of $G$ is well-defined so that $F(a)-F(0)=G(a)+G(0)$.
\end{enumerate}

We can extend the first property for the countably many base functions, as following.

\textbf{Property.} Fix the shifting value $a$. For the base functions $f_i$ with shifting functions $g_i$, if $f=\sum_i f_i$ and $g=\sum_i g_i$ has the radius of convergence $r_f$ and $r_g$ respectively, then $f,g,a$ satisfies the shifting property at $x\in (-r,r)$, where $r=\min(r_f,r_g)$.

Now suppose that $f(x)=1$, then $g(x)=0$ satisfies the shifting property. Integrating both side gives $f(x)=x$ and $g(x)=c_1$, and to satisfy the shifting property, we need $c_1=a/2$. Integration gives $f(x)=x^2$ and $g(x)=ax+c_2$, where $c_2=0$. This goes on. Then we get the following result, which gives the whole data of shifting functions of base function $f(x)=x^n$. We can let $a=1$ here and retrieve $a$ value under the right order relative to $x$. 

\textbf{Shifting functions for polynomials.} Let the base function is $f(x)=x^n$. Then the shifting function $g(x)$ is the polynomial with order $n-1$, which has only nonzero even terms if $n$ is odd and only nonzero odd terms if $n$ is even. If $n=2k$ for some natural number $k$, then $g(x)$ contains the linear term $x$, with coefficients $-G_{2k}$. Here, $G_{2k}$ is the sequence of Genocchi numbers of even index, which can be generated by
\begin{equation}
t\tan\left(\frac{t}{2}\right)=\sum(-1)^n G_{2n} \frac{t^{2n}}{(2n)!}.
\end{equation}
This can be obtained from
\begin{equation}
\frac{2t}{e^t+1}=\sum_{n=1}^\infty G_n\frac{t^n}{n!}
\end{equation}
and changing the variable $t$ to $it$\cite{Com70}.
This gives that for $f_{2k}(x)=x^{2k}$ then
\begin{equation}
g_{2k}(x)=-\sum_{i=1}^k G_{2k-2i+2}\frac{(2k)!}{(2i-1)!(2k-2i+2)!}{x^{2i-1}}.
\end{equation}
and for $f_{2k-1}(x)=x^{2k-1}$,
\begin{equation}
g_{2k-1}(x)=\frac{1}{2k}\frac{dg_{2k}(x)}{dx}.
\end{equation}

Some of the first shifting functions are
\begin{enumerate}
\item $g(x)=1/2$ for $f(x)=x$
\item $g(x)=x$ for $f(x)=x^2$
\item $g(x)=\frac{3}{2}x^2-\frac{1}{4}$ for $f(x)=x^3$
\item $g(x)=2x^3-x$ for $f(x)=x^4$
\item $g(x)=\frac{5}{2}x^4-\frac{5}{2}x^2+\frac{1}{2}$ for $f(x)=x^5$
\item $g(x)=3x^5-5x^3+3x$ for $f(x)=x^6$
\end{enumerate}
and so on.

This result can be used to calculate the shifting function of $\cos(x)$. Notice that we have the power series expression
\begin{equation}
\cos(x)=\sum_{i=0}^\infty (-1)^i \frac{x^{2i}}{(2i)!}.
\end{equation}
Then using the above polynomials, we can obtain the shifting function as
\begin{equation}
g(x)=-\sum_{i=1}^{\infty}\sum_{j=1}^{i} (-1)^i\frac{G_{2i-2j+2}}{(2j-1)!(2i-2j+2)!}x^{2j-1}.
\end{equation}
Substituting $i-j+1$ to $k$ gives
\begin{equation}
g(x)=\sum_{k=1}^{\infty}(-1)^k\frac{G_{2k}}{(2k)!}\sum_{j=1}^{\infty} (-1)^j\frac{1}{(2j-1)!}x^{2j-1},
\end{equation}
which gives
\begin{equation}
g(x)=\tan\left(\frac{1}{2}\right)\sin(x),
\end{equation}
which is the desired result, due to the relation between $t\tan(t/2)|_{t=1}$ and $G_{2k}$.

Thus we have one strategy to find out the shifting function for any base function and shifting value. What we need to do is to represent the base function as power series form, change each polynomial terms into the shifting functions as given above, and take the whole sum and calculate the function.

\section{Appendix: Generalized spin susceptibility when $\delta\neq 0$}



In this section, the spin susceptibility in the main text is generalized to the case away from the fine tuning $s_R \neq s_D$. We quantify a small deviation $\delta$ between the SOC terms $s_R=s+\delta$ and $s_D=s-\delta$. 
Since the the emergent SU(2) symmetry is lost when 
$\delta \neq 0$, the eigenvalue $\sigma^z$ is not good quantum number anymore. Although the spin susceptibility $\chi^{\mu}_{\delta=0}(\vec{q})$ is not appropriate to signal the magnetic instability, the nesting is expected not to be damped so much. To estimate the magnetic instability, we go back to the linear response theory and generalize the formula to $\chi^{\mu}_{\delta\neq 0}(\vec{q})$.
\begin{figure}[t]
\subfloat[]{\label{fig:dsuscepqx}\includegraphics[width=0.24\textwidth]{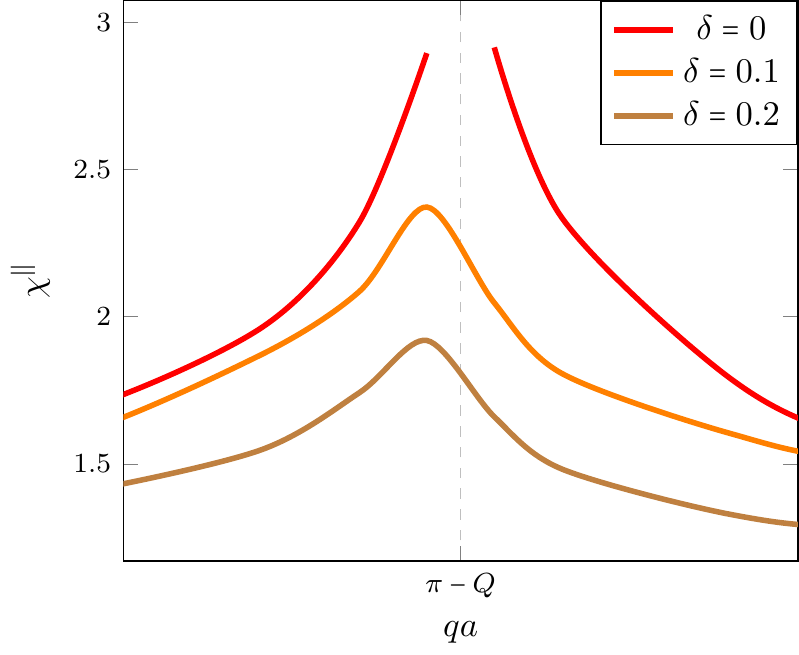}}
\subfloat[]{\label{fig:dsuscepqz}\includegraphics[width=0.24\textwidth]{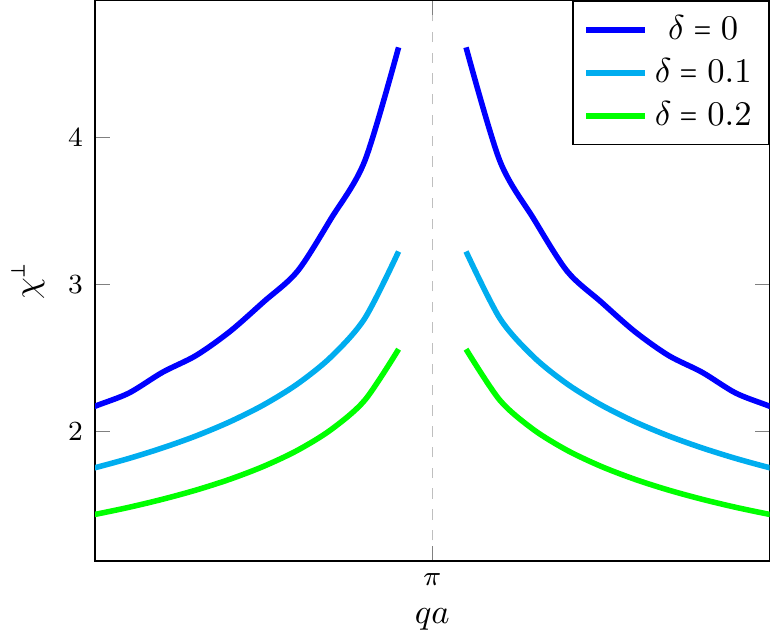}}
\caption{The plot $\chi^{\parallel,\perp}(\vec{q}=(q,q))$ for the half-filled square lattice with $\vec{Q}=(Q,Q)=(0.6\pi, 0.6\pi)$, and $\delta=|s_R-s_D|=0, 0.1, 0.2$. For the numerical evaluation, the susceptibility is regularized to include small $i\epsilon$ in the denominator, which enables us to deal with the divergence at the peak. As $\delta$ increases, the inability to magnetize the system decreases idly.}
\label{fig:dsuscepq}
\end{figure}
Repeating the Fourier transform Eq.\eqref{eqn:hams},
\begin{equation}
\begin{split}
H_{\textrm{SOC}}&=\sum_{\vec{k},\alpha,\beta}c_{\vec{k}\alpha}^\dagger c_{\vec{k}\beta}\left(2\sin{k_x}(s_{\textrm{D}}\sigma_{\alpha\beta}^x-s_{\textrm{R}} \sigma_{\alpha\beta}^y)\right.\\
&\left.+2\sin{k_y}(s_{\textrm{R}}\sigma_{\alpha\beta}^x-s_{\textrm{D}}\sigma_{\alpha\beta}^y)\right)\\
=\sum_{\vec{k}}&\Big[\sqrt{2}s{\sigma}^z(\text{sin}k_x+\text{sin}k_y)
+\sqrt{2}\delta{\sigma}^x(\text{sin}k_x-\text{sin}k_y)\Big]
\end{split}
\label{eq:SOCdelta}
\end{equation} 
where we rotated the spin space $(\sigma^x-\sigma^y)/\sqrt{2} \rightarrow {\sigma}^z$, $-(\sigma^x+\sigma^y)/\sqrt{2} \rightarrow {\sigma}^x$ in the second step.
In case $\delta \neq 0$, ${\sigma}^z$ is not good quantum number, rather the spin axis diagonalizing Eq.(\ref{eq:SOCdelta}) rotates in $\vec{k}$-space. Denoting the coefficients $A_{\vec{k}}=\sqrt{2}(\text{sin}k_x+\text{sin}k_y)$ and $B_{\vec{k}}=\sqrt{2}(\text{sin}k_x-\text{sin}k_y)$, Eq.(\ref{eq:SOCdelta}) is diagonalized as
\begin{equation}
E_{\tilde{\sigma}^z}(\vec{k})=\epsilon_{\vec{k}}+\tilde{\sigma}^z(\theta_{\vec{k}})\sqrt{s^2A_{\vec{k}}^2+\delta^2B_{\vec{k}}^2}
\label{eq:SOCdiag}
\end{equation}
where $\epsilon_{\vec{k}}$ is the non-interacting dispersion and $\tilde{\sigma}^z=\sigma^z\text{cos}(\theta_{\vec{k}})+\sigma^x\text{sin}(\theta_{\vec{k}})$ and $\theta_{\vec{k}}=\text{tan}^{-1}(\delta B_{\vec{k}}/sA_{\vec{k}})$. The ground state is filled with the states below the chemical potential, $\vert\text{GS} \rangle=\prod_{E_{\tilde{\sigma}}<0}\tilde{c}^{\dagger}_{\vec{k}\tilde{\sigma}}\vert 0\rangle$ where $\tilde{c}^{\dagger}_{\vec{k}\uparrow}=\text{cos}{\frac{\theta_{\vec{k}}}{2}}c^{\dagger}_{\vec{k}\uparrow}+\text{sin}{\frac{\theta_{\vec{k}}}{2}}c^{\dagger}_{\vec{k}\downarrow}$ and
$\tilde{c}^{\dagger}_{\vec{k}\downarrow}=-\text{sin}{\frac{\theta_{\vec{k}}}{2}}c^{\dagger}_{\vec{k}\uparrow}+\text{cos}{\frac{\theta_{\vec{k}}}{2}}c^{\dagger}_{\vec{k}\downarrow}$. This transformation between the creation/annihilation operators can be rewritten as
\begin{eqnarray}
c^{\dagger}_{\vec{k},\uparrow}&=\langle\vec{k}\uparrow|\tilde{\uparrow}(\vec{k})\rangle\tilde{c}^{\dagger}_{\vec{k},\uparrow}+\langle\vec{k}\uparrow|\tilde{\downarrow}(\vec{k})\rangle\tilde{c}^{\dagger}_{\vec{k},\downarrow}
\nonumber\\
c^{\dagger}_{\vec{k},\downarrow}&=\langle\vec{k}\downarrow|\tilde{\uparrow}(\vec{k})\rangle\tilde{c}^{\dagger}_{\vec{k},\uparrow}+\langle\vec{k}\downarrow|\tilde{\downarrow}(\vec{k})\rangle\tilde{c}^{\dagger}_{\vec{k},\downarrow}
\label{eq:Crea/anni}
\end{eqnarray}
where $|\tilde{\sigma}(\vec{k})\rangle=\tilde{c}^{\dagger}_{\vec{k},\tilde{\sigma}}|0\rangle$ and $|\vec{k},\sigma\rangle=c^{\dagger}_{\vec{k},\sigma}|0\rangle$.

Now we introduce the text field $H'=-2h(S^x_{\vec{q}}+S^x_{-\vec{q}})$ to measure the response. The ground state is perturbed as
\begin{eqnarray}
|\text{GS}'\rangle=|\text{GS}\rangle +h\sum_{\vec{k},\sigma,\tilde{\sigma}_1,\tilde{\sigma}_2}&&\frac{\langle \text{GS}|\tilde{c}^{\dagger}_{\vec{k}\tilde{\sigma}_2}\tilde{c}_{\vec{k}\pm\vec{q},\tilde{\sigma}_1}c^{\dagger}_{\vec{k}\pm\vec{q}\sigma}c_{\vec{k}-\sigma}|\text{GS}\rangle}{E_{\vec{k}\pm\vec{q},\tilde{\sigma}_1}-E_{\vec{k},\tilde{\sigma}_2}}
\nonumber\\
&&\times\tilde{c}^\dagger_{\vec{k}\pm\vec{q},\tilde{\sigma}_1}\tilde{c}_{\vec{k},\tilde{\sigma}_2}|\textrm{GS}\rangle
\label{eq:pertGS}
\end{eqnarray}
and the susceptibility is
\begin{eqnarray}
\chi_{\delta}^x(\vec{q})=\frac{\text{d}}{\text{d}h}&\langle  \text{GS}'\vert S^x_{\vec{q}}&\vert \text{GS}'\rangle \Big\vert_{h\rightarrow 0}
\nonumber\\
=\sum_{\vec{k},\tilde{\sigma}_1,\tilde{\sigma}_2}|\langle & \tilde{\sigma}_1(\vec{k}+\vec{q})&|\sigma^x_{\vec{q}}|\tilde{\sigma}_2(\vec{k})\rangle |^2\frac{f_{\vec{k},\tilde{\sigma}_2}-{f_{\vec{k}+\vec{q},\tilde{\sigma}_1}}}{E_{\vec{k}+\vec{q},\tilde{\sigma}_1}-E_{\vec{k},\tilde{\sigma}_2}}\quad\quad\quad
\label{eq:GeneSusc}
\end{eqnarray}
Here, we used Eq.(\ref{eq:Crea/anni}) and
\begin{equation}
\begin{split}
&|\langle \text{GS}|\sum_{\sigma}\Big(c^{\dagger}_{\vec{k}+\vec{q},\sigma}c_{\vec{k},-\sigma}\Big)\tilde{c}^\dagger_{\vec{k}+\vec{q},\tilde{\sigma}_1}\tilde{c}_{\vec{k},\tilde{\sigma}_2}|\text{GS}\rangle|^2 \\
&=|\sum_{\sigma}\langle \tilde{\sigma}_1(\vec{k}+\vec{q})|\vec{k}+\vec{q},\sigma\rangle\langle \vec{k},-\sigma |\tilde{\sigma}_2(\vec{k} \rangle |^2(f_{\vec{k},\tilde{\sigma}_2}-{f_{\vec{k}+\vec{q},\tilde{\sigma}_1}})\\
&=|\langle \tilde{\sigma}_1(\vec{k}+\vec{q})|\sigma^x_{\vec{q}}|\tilde{\sigma}_2(\vec{k})\rangle |^2(f_{\vec{k},\tilde{\sigma}_2}-{f_{\vec{k}+\vec{q},\tilde{\sigma}_1}})
\end{split}
\label{eq:Susfactor}
\end{equation}

The first factor in the last line Eq.(\ref{eq:GeneSusc}) accounts for the probability of nesting.
At the fine tuning $\delta=0$, $|\langle \tilde{\sigma}_1(\vec{k}+\vec{q})|\sigma^x_{\vec{q}}|\tilde{\sigma}_2(\vec{k})\rangle |^2=\delta_{\sigma_1,-\sigma_2}$ and $\chi^x_{\delta=0}(\vec{q})$ agrees with the formula in the main text. However, this additional factor is less than 1 and depress the overall weight for generic $\delta\neq 0$. Figure \ref{fig:dsuscepq} is the numerical calculation of the susceptibilies in half-filled square lattice, drawn near peak, which shows the depressed peaks as $\delta$ increases. Eq.(\ref{eq:GeneSusc}) can be  directly generalized to other directions $\mu=y,z$ by replacing $\sigma^x_{\vec{q}}$ to $\sigma^{y,z}_{\vec{q}}$.

\vskip 1cm


\begin{acknowledgments}

\noindent	
{\em Acknowledgments.---}
H.-J. K. P, H.J.Y and S.B.L. is supported by the KAIST startup and National Research Foundation Grant (NRF-2017R1A2B4008097). 
\end{acknowledgments}



\end{document}